\documentclass[conference,letterpaper,twoside,twocolumn]{IEEEtran}
\renewcommand{\IEEEQED}{\IEEEQEDopen} % IEEE Trans. IT uses open squares
\usepackage{amsmath,amssymb,amscd,latexsym,dsfont,mathtools}
\usepackage{float}
\usepackage{color}
\usepackage{graphicx}
\usepackage{comment}
\usepackage{subfigure}
\usepackage{enumerate}
\usepackage{stfloats}
\usepackage{psfrag}
\usepackage{setspace}
\usepackage{cite}
\usepackage{balance}
\usepackage{multirow}
\usepackage[para]{threeparttable}
\usepackage{pstricks}
\IEEEoverridecommandlockouts

\renewcommand{\markboth}[1]{\renewcommand{\leftmark}{#1}\renewcommand{\rightmark}{#1}}
\newcommand{\argmin}[2]{\mathop{\mathrm{argmin}}_{#1} {#2}}
\newcommand{\qfunc}[0]{\mathrm{Q}}
\newcommand{\refl}[0]{\mathrm{refl}}
\newcommand{\inv}[0]{\mathrm{inv}}
\newcommand{\trans}{\mathsf{T}}

\newcommand{\SNR}[0]{{\rho}}
\newcommand{\Es}{E_{\mathrm{s}}}

\newtheorem{theorem}{Theorem}

\newtheorem{remark}{Remark}

\newcommand{\PC}{P_{\C}}

\newcommand{\Pt}{P}

\newcommand{\Pj}{P_{j}}

\newcommand{\C}{\mathbb{C}}
\newcommand{\X}{\mathbb{X}}

\newcommand{\xx}{\boldsymbol{x}}
\newcommand{\bb}{\boldsymbol{b}}
\newcommand{\cc}{\boldsymbol{c}}

\newcommand{\pp}{\boldsymbol{p}}

\newcommand{\aalpha}{\boldsymbol{\alpha}}
\renewcommand{\aa}{\boldsymbol{a}}
\newcommand{\setX}{\mathcal{X}}
\newcommand{\setR}{\mathcal{R}}
\newcommand{\setW}{\mathcal{W}}

\newcommand{\tab}{\,\,\,}
\newcommand{\tabb}{\,\,}

\newcommand{\eqsref}[2]{(\ref{#1})--(\ref{#2})}
\newcommand{\figref}[1]{Fig.~\ref{#1}}

\newcommand{\tabref}[1]{Table~\ref{#1}}

\newcommand{\theref}[1]{Theorem~\ref{#1}}

\title{General BER Expression for One-Dimensional Constellations}
\author{%
\IEEEauthorblockN{Mikhail Ivanov, Fredrik Br\"{a}nnstr\"{o}m, Alex Alvarado{\IEEEauthorrefmark{4}}, and Erik Agrell\\}
\IEEEauthorblockA{Department of Signals and  Systems, Chalmers  University of Technology, Gothenburg, Sweden\\} \IEEEauthorblockA{ \IEEEauthorrefmark{4}Department of Engineering, University of Cambridge, UK\\ \emph{\{mikhail.ivanov,fredrik.brannstrom,agrell\}@chalmers.se, alex.alvarado@ieee.org}
}
\thanks{Research supported by The British Academy and The Royal Society (via the Newton International Fellowship scheme), UK, and by the Swedish Research Council, Sweden (under grant \#621-2006-4872 and \#621-2011-5950).} }%

\begin{document}
\maketitle
%\tableofcontents

%%%%%%%%%%%%%%%%%%%%%%%%%%%%%%%%%%%%%%%%%%%%%%%%%%%%%%%%%%%%%%%%%%%%%%%%%%%%%%%%%%%%%%%%%%%%%%%%%%%%%%%%%%%%%%%%%%%%%%%%%%%%%%%%%%%%%%%%%%%%
%%%%%%%%%%%%%%%%%%%%%%%%%%%%%%%%%%%%%% Abstract %%%%%%%%%%%%%%%%%%%%%%%%%%%%%%%%%%%%%%%%%%%%%%%%%%%%%%%%%%%%%%%%%%%%%%%%%%%%%%%%%%%%%%%%%%%%
%%%%%%%%%%%%%%%%%%%%%%%%%%%%%%%%%%%%%%%%%%%%%%%%%%%%%%%%%%%%%%%%%%%%%%%%%%%%%%%%%%%%%%%%%%%%%%%%%%%%%%%%%%%%%%%%%%%%%%%%%%%%%%%%%%%%%%%%%%%%
\begin{abstract}
A novel general ready-to-use bit-error rate (BER) expression for one-dimensional constellations is developed. The BER analysis is performed for bit patterns that form a labeling. The number of patterns for equally spaced $M$-PAM constellations with different BER is analyzed.
\end{abstract}

%%%%%%%%%%%%%%%%%%%%%%%%%%%%%%%%%%%%%%%%%%%%%%%%%%%%%%%%%%%%%%%%%%%%%%%%%%%%%%%%%%%%%%%%%%%%%%%%%%%%%%%%%%%%%%%%%%%%%%%%%%%%%%%%%%%%%%%%%%%%
%%%%%%%%%%%%%%%%%%%%%%%%%%%%%%%%%%%%%% Introduction %%%%%%%%%%%%%%%%%%%%%%%%%%%%%%%%%%%%%%%%%%%%%%%%%%%%%%%%%%%%%%%%%%%%%%%%%%%%%%%%%%%%%%%%
%%%%%%%%%%%%%%%%%%%%%%%%%%%%%%%%%%%%%%%%%%%%%%%%%%%%%%%%%%%%%%%%%%%%%%%%%%%%%%%%%%%%%%%%%%%%%%%%%%%%%%%%%%%%%%%%%%%%%%%%%%%%%%%%%%%%%%%%%%%%
\section{Introduction and Motivation}\label{sec:intro}

Current wireless communication systems are based on the bit-interleaved coded modulation (BICM) paradigm introduced in \cite{Zehavi92may} and later studied in \cite{Caire98,Fabregas08_Book}. One key element in these systems is the calculation of logarithmic likelihood ratios (LLR, also known as L-values) for the received bits, which are passed to the channel decoder. The coded performance analysis of such systems is generally not straightforward, and is usually carried out either numerically by Monte-Carlo simulation, or in terms of lower and upper bounds \cite[Sec.~4]{Caire98}, \cite[Ch.~4]{Fabregas08_Book}. The calculation of LLRs is crucial also in many other coded systems. In this paper, we analyze the \emph{uncoded} performance over the additive white Gaussian noise (AWGN) channel.

A symbol-based demodulator (SD) is the most natural way of decoding symbols transmitted through the channel. This approach is optimal in terms of symbol-error rate (SER). The bit-error rate (BER) performance of the SD is well documented in literature, e.g.~\cite[Ch.~5]{Proakis00_Book}, \cite[Ch.~10]{Simon95_Book}, \cite{ChoYoon02jul,Lee86may,Lassing03nov,Agrell04dec,Lassing03b,Szczecinski06b} and references therein.
On the other hand, in a coded system, such as BICM, soft or hard information on the received \emph{bits} is passed to the decoder, and thus, bit-wise decisions are more relevant than symbol-wise decisions. The optimal \emph{bit-wise} demodulator (BD) minimizing the BER implies the calculation of (exact) L-values for the received bits. The uncoded performance of such a demodulator has been studied in \cite{Simon05feb}, where closed-form expressions for the BER for 4-ary pulse amplitude modulation (PAM) with the binary reflected Gray code (BRGC) \cite{Gray53,Agrell04dec,Agrell07} are presented. Due to the complexity of the BD, the calculation of L-values in practical systems is usually done based on the so-called max-log approximation \cite[eq.~(5)]{Viterbi98feb}, \cite[eq.~(1)]{Tsgr1_15_1093}. We call this demodulator the approximate BD (ABD). The three above demodulators (SD, BD, and ABD) have been recently numerically compared from a mutual information point of view in \cite{Fertl12} for multiple-input multiple-output BICM systems.

In this paper, we prove the equivalence of the SD and the ABD in terms of uncoded BER for any constellation and labeling. Due to this equivalency, we go on and study the ABD for one-dimensional constellations. To this end, we introduce a novel ready-to-use BER expression valid for any one-dimensional constellation and binary labeling. The analysis is performed for bit-patterns that form a labeling.

%%%%%%%%%%%%%%%%%%%%%%%%%%%%%%%%%%%%%%%%%%%%%%%%%%%%%%%%%%%%%%%%%%%%%%%%%%%%%%%%%%%%%%%%%%%%%%%%%%%%%%%%%%%%%%%%%%%%%%%%%%%%%%%%%%%%%%%%%%%%
%%%%%%%%%%%%%%%%%%%%%%%%%%%%%%%%%%%%%% System model %%%%%%%%%%%%%%%%%%%%%%%%%%%%%%%%%%%%%%%%%%%%%%%%%%%%%%%%%%%%%%%%%%%%%%%%%%%%%%%%%%%%%%%%
%%%%%%%%%%%%%%%%%%%%%%%%%%%%%%%%%%%%%%%%%%%%%%%%%%%%%%%%%%%%%%%%%%%%%%%%%%%%%%%%%%%%%%%%%%%%%%%%%%%%%%%%%%%%%%%%%%%%%%%%%%%%%%%%%%%%%%%%%%%%
\section{Preliminaries}\label{sec:syst_mod}

\subsection{Notation Convention}

The following notation is used throughout the paper. Lowercase letters $x$ denote real scalars and boldface letters $\xx$ denote a row vector of scalars. Blackboard bold letters $\X$ denote matrices with elements $x_{i,j}$ in the $i$th row and the $j$th column and $(\cdot)^\trans$ denotes transposition. Calligraphic capital letters $\setX$ denote sets, where the set of real numbers is denoted by $\setR$. The  binary complement of $x \in \{0, 1\}$ is denoted by $\bar{x}=1-x$. Binary addition (exclusive-OR) of two bits $a$ and $b$ is denoted by $a\oplus b$. Random variables are denoted by capital letters $X$ and probabilities by $\Pr\{\cdot\}$. The Gaussian Q-function is defined as $\qfunc(x) \triangleq \frac{1}{\sqrt{2\pi}}\int_{x}^{\infty}\mathrm{e}^{-\frac{t^2}{2}}\,\mathrm{d}t$.

\subsection{System Model}

In this paper we analyze a system where a vector of binary data $\bb = [b_1,\dots,b_m]$ is fed to a modulator. The modulator carries out a one-to-one mapping from $\bb$ to one of the $M$ constellation points $x \in \setX = \{s_1,\dots, s_M\}$, where $s_1<s_2<\ldots<s_M$, for transmission over the physical channel, where $M = 2^m$. The modulator is defined as the function $\Phi: \{0, 1\}^m \rightarrow \setX$ with a corresponding inverse function $\Phi^{-1}:\setX \rightarrow \{0,  1\}^m$.

For PAM constellations, $s_i = -d(M-2i+1), i=1,\dots,M$, where $d = \sqrt{{3}/{(M^2-1)}}$ to normalize the constellation to unit average energy, i.e., $\Es = \frac{1}{M}\sum_{i=1}^M{s_i^2} = 1$. We assume the bits to be independent and identically distributed (i.i.d.) with $\Pr\{B_j = u\} = 0.5$,\,$\forall j$ and $u \in \{0,1\}$, and thus, the symbols are equiprobable, i.e., $\Pr\{X = s_i\} = 1/M$,\, $\forall i$.

The modulator is defined by the constellation and its binary labeling. A binary labeling is specified by the matrix $\C = [\cc_1^\trans, \dots, \cc_M^\trans]^\trans$ of dimensions $M$ by $m$, where the $i$th row $\cc_i = [c_{i,1},\dots, c_{i,m}]$ is the binary label of the constellation point $s_i$, i.e., $\Phi(\cc_i)=s_i$.

In this paper we consider a discrete time memoryless AWGN channel with output $y = x + \eta$, where $x\in \setX$ and the noise sample $\eta$ is a zero-mean Gaussian random variable with variance $N_0/2$. The conditional PDF of the channel output is given by
\begin{equation}
    p_{Y|X}(y|x) = \sqrt{\frac{\SNR}{\pi}}\mathrm{e}^{-\SNR(y - x)^2},
    \label{eq:gauss}
\end{equation}
where the average signal to noise ratio (SNR) is defined as $\SNR \triangleq {\Es}/{N_0} = {1}/{N_0}$.

The observation $y$ is used by the demodulator to decide on the transmitted binary sequence, i.e., to produce $\hat{\bb} = [\hat{b}_1, \dots, \hat{b}_m]$.

%%%%%%%%%%%%%%%%%%%%%%%%%%%%%%%%%%%%%%%%%%%%%%%%%%%%%%%%%%%%%%%%%%%%%%%%%%%%%%%%%%%%%%%%%%%%%%%%%%%%%%%%%%%%%%%%%%%%%%%%%%%%%%%%%%%%%%%%%%%%
%%%%%%%%%%%%%%%%%%%%%%%%%%%%%%%%%%%%%% Demods %%%%% %%%%%%%%%%%%%%%%%%%%%%%%%%%%%%%%%%%%%%%%%%%%%%%%%%%%%%%%%%%%%%%%%%%%%%%%%%%%%%%%%%%%%%%%
%%%%%%%%%%%%%%%%%%%%%%%%%%%%%%%%%%%%%%%%%%%%%%%%%%%%%%%%%%%%%%%%%%%%%%%%%%%%%%%%%%%%%%%%%%%%%%%%%%%%%%%%%%%%%%%%%%%%%%%%%%%%%%%%%%%%%%%%%%%%
\subsection{Demodulators}

The SD makes a hard decision on the transmitted symbol and returns the length-$m$ binary label of that symbol, i.e.,
\begin{equation}
   \hat{\bb}^{\mathrm{SD}} \triangleq \mathrm{\Phi}^{-1}\left(\argmin{x \in \setX}{(y-x)^2}\right).
\label{eq:SD}
\end{equation}
The SD in~\eqref{eq:SD} is optimal in terms of minimizing the SER, but it does not necessarily minimize the BER.

To minimize the BER the optimal BD should be used. The BD calculates (a posteriori) L-values for the $m$ bits based on the observation $y$, i.e.,
\begin{align}
\label{eq:llr_almost}
    l_j(y) &\triangleq \log{\frac{\Pr\{B_j = 1|Y=y\}}{\Pr\{B_j = 0|Y=y\}}} \\
    &= \log{ \frac{\sum_{x \in \setX_{j, 1}}{\mathrm{e}^{-\SNR(y-x)^2}}}{\sum_{x \in \setX_{j, 0}}{\mathrm{e}^{-\SNR(y-x)^2}}}},
    \label{eq:llr}
\end{align}
 where $j=1,\dots,m$ and  $\setX_{j,u} \triangleq \{s_i \in \setX: c_{i,j} = u,\, \forall i \}$. To pass from~\eqref{eq:llr_almost} to \eqref{eq:llr} Bayes' rule was used together with the i.i.d.~assumption of the bits and the conditional PDF in \eqref{eq:gauss}.

The implementation of the BD in its exact form~\eqref{eq:llr} is complicated, especially for large constellations, as it requires calculation of the logarithm of a sum of exponentials. To overcome this problem, approximations are usually used in practice. The most common approximation is the so-called max-log approximation ($\log{\sum_i \mathrm{e}^{\lambda_i}} \approx \max_i{\lambda_i}$) \cite[eq.~(3.2)]{Zehavi92may}, \cite[eq.~(9)]{Caire98}, \cite[eq.~(5)]{Viterbi98feb}, \cite[eq.~(8)]{Robertson95jun}, which used in~\eqref{eq:llr} gives
\begin{equation}
    \tilde{l}_j(y) = \SNR\left[\min_{x \in \setX_{j, 0}}{(y-x)^2} - \min_{x \in \setX_{j, 1}}{(y-x)^2}\right].
    \label{eq:max_log_rrl}
\end{equation}
The ABD is defined as the demodulator that applies the following decision rule
\begin{equation}
    \hat{b}_j^{\mathrm{ABD}} =
    \begin{cases}
    1 &\text{if } \tilde{l}_j(y) \ge 0,\\
    0 & \text{otherwise}.
    \end{cases}
\label{eq:decision_rule}
\end{equation}

The next theorem gives proof for the equivalence of the SD and the ABD. This was mentioned in \cite[Sec.~IV-A]{Fertl12}, however, no proof was given there.
\begin{theorem}\label{theor:equiv}
    For any $\SNR$, $\setX$, and $\C$, $\hat{b}_j^{\mathrm{SD}} = \hat{b}_j^{\mathrm{ABD}}$ for all $j=1,\dots,m$.
\end{theorem}
\begin{IEEEproof}
Combining \eqref{eq:decision_rule} and \eqref{eq:max_log_rrl}, the decision rule for the ABD can be written as
\begin{equation*}
    \hat{b}_j^{\mathrm{ABD}} =
    \begin{cases}
        1, & \min_{x\in \setX_{j,  0}}{(y-x)^2} \ge \min_{x\in \setX_{j,  1}}{(y-x)^2},\\
        0, & \min_{x \in \setX_{j,  0}}{(y-x)^2} < \min_{x\in \setX_{j,  1}}{(y-x)^2},
    \end{cases}
\end{equation*}
which can be simplified to
\begin{equation}
    \hat{b}_j^{\mathrm{ABD}} = \argmin{ u\in \{0,  1\}}{\left\{\min_{x \in \setX_{j, u}}{(y-x)^2}\right\}}.
\label{eq:equiv_proof2}
\end{equation}
Since $\min_{u\in \{0,  1\}}{\left\{\min_{x \in \setX_{j, u}}{(y-x)^2}\right\}} = \min_{x \in \setX}{(y-x)^2}$ for any $\setX$, $\SNR$, and $\C$, the symbol found by the ABD in~\eqref{eq:equiv_proof2} will always be the closest $x \in \setX$ to $y$ in terms of Euclidean distance (ED), regardless of the bit position $j$. This is the same rule used in \eqref{eq:SD}, which completes the proof.
\end{IEEEproof}

\theref{theor:equiv} states that the SD and the ABD are equivalent and optimal in terms of minimizing the SER for any constellation\footnote{The proof of \theref{theor:equiv} was given for one-dimensional constellations only, however, its extension to any multi-dimensional constellation is straightforward.} and any labeling. Because of this, from now on we only consider the ABD.

%%%%%%%%%%%%%%%%%%%%%%%%%%%%%%%%%%%%%%%%%%%%%%%%%%%%%%%%%%%%%%%%%%%%%%%%%%%%%%%%%%%%%%%%%%%%%%%%%%%%%%%%%%%%%%%%%%%%%%%%%%%%%%%%%%%%%%%%%%%%
%%%%%%%%%%%%%%%%%%%%%%%%%%%%%%%%%%%%%% General BER %%%%%%%%%%%%%%%%%%%%%%%%%%%%%%%%%%%%%%%%%%%%%%%%%%%%%%%%%%%%%%%%%%%%%%%%%%%%%%%%%%%%%%%%%
%%%%%%%%%%%%%%%%%%%%%%%%%%%%%%%%%%%%%%%%%%%%%%%%%%%%%%%%%%%%%%%%%%%%%%%%%%%%%%%%%%%%%%%%%%%%%%%%%%%%%%%%%%%%%%%%%%%%%%%%%%%%%%%%%%%%%%%%%%%%
\section{BER for One-Dimensional Constellations}\label{sec:patterns}
The BER for a given labeling $\C$ can be expressed as
\begin{equation}
\PC 	= \frac{1}{m}\sum_{j=1}^m \Pj, \label{eq:P_C1.0}
\end{equation}
where the BER for the $j$th bit position $\Pj \triangleq\Pr\{\hat{B}_j \neq b_j|B_j=b_j\}$ can be written as
\begin{equation}
P_j = \frac{1}{M}\sum_{i = 1}^{M}{\Pr\{\hat{B}_j \neq c_{i,j}  | X = s_i\}}
\label{eq:P_C1}
\end{equation}
using the law of total probability. The BER for the $j$th bit position $P_j$ depends only on the subconstellations $\setX_{j,0}$ and $\setX_{j,1}$ (cf.~\eqref{eq:llr}--\eqref{eq:max_log_rrl}), i.e., on the $j$th column of $\C$, such that $P_j = P([c_{1,j},\dots,c_{M,j}])$.

We define a bit pattern (or simply pattern) as a length-$M$ binary vector $\pp = [p_1,\dots,p_M] \in \{0,1\}^M$ with Hamming weight $M/2$. The labeling $\C$ can now be defined by $m$ patterns, each corresponding to one column of $\C$. We index the patterns as $\pp_w$ with $w$ being the decimal representation of the vector $\pp$, i.e., $w = \sum_{i = 1}^{M}{2^{M-i}p_i}$. For example, for $M=4$, the pattern $[0,1,0,1]$ is indexed as $\pp_5$ (cf.~\tabref{tab:all_seq}). The BER for the labeling $\C$ does not depend on the order of its columns, and thus, the BER for the labeling $\C$ is fully determined by a set of $m$ patterns $\setW= \{w_1,\dots,w_m\}$.

Based on the previous discussion, from now on we concentrate our analysis only on patterns (and not on labelings), i.e., on the function $P(\pp)$, however, to simplify the notation, the dependency on the pattern will be omitted.

To analyze the BER of a pattern (PBER), the observation space $\setR$ is split into two disjoint decision regions, i.e., $\Gamma_{0} = \{y \in \setR: \hat{b} = 0\}$ and $\Gamma_{1} = \{y \in \setR: \hat{b} = 1\}$ such that $\Gamma_{0} \cup \Gamma_{1} = \setR$.

Using the definition of $\Gamma_{0}$ and $\Gamma_{1}$, the PBER for the pattern $\pp$ can be rewritten as
\begin{equation}
P = {\frac{1}{M}\sum_{i = 1}^{M}{\Pr\{Y \in \Gamma_{\bar{p}_{i}} | X = s_i\}}}.
\label{eq:P_j}
\end{equation}
By expressing $P$ as in \eqref{eq:P_j}, it is clear that the PBER in \eqref{eq:P_C1} can be calculated using the decision regions $\Gamma_{0}$ and $\Gamma_{1}$ only, as opposed to alternative approaches where \eqref{eq:P_j} is expressed in terms of the PDF of the L-values (cf.~\cite[eq.~(19)]{Benjillali07}, \cite[Sec.~IV]{Alvarado07d}).

Decision thresholds (or simply thresholds), denoted by $\beta_k$, where $k = 1,2,\dots$ stands for the index of the threshold, are defined as the points that separate the decision regions for zeros and ones, and thus, they fully determine the PBER in~\eqref{eq:P_j}. The thresholds for the ABD are the midpoints between the constellation points labeled with different bits, which follows directly from~\eqref{eq:equiv_proof2}.

The BER expression for the ABD and an M-PAM constellation with any labeling is well known and can be found in~\cite[eq.~(21)]{Agrell07}. The PBER expression can easily be obtained in a similar way.  In the following theorem, we generalize the result in~\cite[eq.~(21)]{Agrell07} to non-equally spaced constellations and derive a general PBER expression for any one-dimensional constellation.

\begin{theorem}\label{theo:ber_1D_general}
The PBER for the ABD using an arbitrary one-dimensional constellation with a pattern $\pp$ can be expressed~as
\begin{align}
    P &= \frac{1}{2} + \frac{1}{M}\sum_{i = 1}^{M}\sum_{k = 1}^{M-1}g_{i,k}\qfunc\left((\beta_k-s_i)\sqrt{2\SNR}\right),\label{eq:BER_general}
\end{align}
where $\beta_k = \frac{s_k + s_{k+1}}{2}, \tab k=1,\dots,M-1$ and $g_{i,k} \in \{0,\pm1\}$~is
\begin{equation}
g_{i,k} \triangleq (p_{k+1}-p_k)(1-2p_i).\label{eq:g_ik}
\end{equation}
\end{theorem}

\begin{IEEEproof}
Let $v_{i,k}$ be the following conditional probabilities
\begin{align}
\nonumber
v_{i,1} 	&\triangleq \Pr\{Y \le \beta_1|X=s_i\} \\
		& = 1-\qfunc\left((\beta_1-s_i)\sqrt{2\SNR}\right),\label{eq:v_ik.1}\\
\nonumber
v_{i,k} 	&\triangleq \Pr\{\beta_{k-1} < Y \le \beta_k|X=s_i\} \\
		&= \qfunc\left((\beta_{k-1}-s_i)\sqrt{2\SNR}\right)-\qfunc\left((\beta_k-s_i)\sqrt{2\SNR}\right),\label{eq:v_ik.2}\\
\nonumber
v_{i,M} 	&\triangleq \Pr\{\beta_{M-1} < Y|X=s_i\} \\
		&= \qfunc\left((\beta_{M-1}-s_i)\sqrt{2\SNR}\right), \label{eq:v_ik.3}
\end{align}
where $i = 1,\dots, M$, $k = 2,\dots,M-1$, and $\beta_k = \frac{s_k + s_{k+1}}{2}$ for $k=1,\ldots,M-1$. The PBER in~\eqref{eq:P_j} can be rewritten as
\begin{align}
P &= {\frac{1}{M}\sum_{i = 1}^{M}{\Pr\{Y \in \Gamma_{\bar{p}_{i}} | X = s_i\}}}\nonumber\\
 &= \frac{1}{M}\sum_{i = 1}^{M}\sum_{k = 1}^{M}e_{i,k}v_{i,k},
\label{eq:P_ev}
\end{align}
where $e_{i,k} \triangleq p_i \oplus p_k\in\{0,1\}$.

Using \eqref{eq:v_ik.1}--\eqref{eq:v_ik.3} the PBER in \eqref{eq:P_ev} can be expressed as
\begin{align}
   P & = \frac{1}{M}\biggl[\sum_{i = 1}^{M}e_{i,1} + \sum_{i = 1}^{M}\sum_{k = 2}^{M}e_{i,k}\qfunc\left((\beta_{k-1}-s_i)\sqrt{2\SNR}\right) \nonumber\\
   	& \qquad\qquad\qquad - \sum_{i = 1}^{M}\sum_{k = 1}^{M-1}e_{i,k}\qfunc\left((\beta_k-s_i)\sqrt{2\SNR}\right)\biggr]\nonumber\\
   & = \frac{1}{2} + \frac{1}{M}\sum_{i = 1}^{M}\sum_{k = 1}^{M-1}(e_{i,k+1}-e_{i,k})\qfunc\left((\beta_k-s_i)\sqrt{2\SNR}\right),
    \label{eq:P_eQ}
\end{align}
where $\sum_{i = 1}^{M}e_{i,1} = \sum_{i = 1}^{M}p_i \oplus p_1 = M/2$ was used.
To obtain the expression in~\eqref{eq:BER_general}, we express $e_{i,k+1}-e_{i,k}$ in \eqref{eq:P_eQ} as
\begin{align}
e_{i,k+1}-e_{i,k} 	&= p_{k+1} \oplus p_{i} - p_{k} \oplus p_{i}\label{eq:g_ik2.1}\\
				& = (p_{k+1}-p_k)(1-2p_i),\label{eq:g_ik2.5}
\end{align}
where the identity $p_i \oplus p_j = p_i\bar{p}_j + \bar{p}_i p_j$ was used together with $\bar{p}_i = 1-p_i$.

The threshold $\beta_k$ between the constellation points labeled with the same bit does not affect the PBER in~\eqref{eq:BER_general} as $g_{i,k} = 0, \,\forall i$ in~\eqref{eq:g_ik}.
\end{IEEEproof}
\begin{remark}\label{rem:ber_general}
\theref{theo:ber_1D_general} gives an expression for the PBER for the ABD. However, \eqref{eq:BER_general} can be used for calculating the PBER when the thresholds $\beta_k$ are not midpoints or, moreover, when they are dependent on the SNR, for example, when the BD is used. Analytical expressions for thresholds for the BD are in general unknown.
\end{remark}

\begin{figure}[t]
\newcommand{\scale}{0.9}
\begin{center}
    \psfrag{SNR}[tc][tc]{$\SNR$~[dB]}
    \psfrag{BER}[bc][Bc][\scale]{}
    \psfrag{(a)}[cc][cc][\scale]{(a)}
    \psfrag{(b)}[cc][cc][\scale]{(b)}
    \psfrag{d12222}[cl][cl][\scale]{$\beta_4$, $\pp_{15}$}
    \psfrag{d2}[cl][cl][\scale]{$\beta_6$, $\pp_{60}$}
    \psfrag{d3}[cl][cl][\scale]{$\beta_7$, $\pp_{102}$}
    \psfrag{d4}[cl][cl][\scale]{$\beta_5$, $\pp_{102}$}
    \psfrag{ s5}[cl][cl][\scale]{$s_5$}
    \psfrag{ s6}[cl][cl][\scale]{$s_6$}
    \psfrag{ s7}[cl][cl][\scale]{$s_7$}
    \psfrag{ s8}[cl][cl][\scale]{$s_8$}
    \psfrag{ b4}[cl][cl][\scale]{}
    \psfrag{ b5}[cl][cl][\scale]{}
    \psfrag{ b6}[cl][cl][\scale]{}
    \psfrag{ b7}[cl][cl][\scale]{}
    \includegraphics[width=0.95\columnwidth]{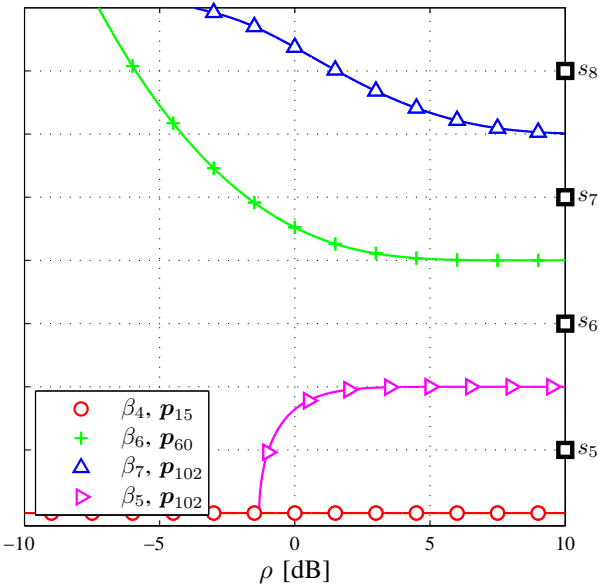}
    \caption{Thresholds  for 8-PAM with different patterns  vs. SNR. Due to the symmetry of the patterns the thresholds are symmetric with respect to zero. Only positive thresholds are shown. Squares represent the constellation points.}
    \label{fig:BD_THR}
\end{center}
\end{figure}

\begin{figure}[t]
\newcommand{\scale}{0.9}
\begin{center}
    \psfrag{SNR}[tc][tc]{$\SNR$~[dB]}
    \psfrag{BER}[bc][Bc][\scale]{BER}
%    \psfrag{d1}[cl][cl][\scale]{$q = 1$}
    \psfrag{d2}[cl][cl][\scale]{$\pp_{60}$}
    \psfrag{d3}[cl][cl][\scale]{$\pp_{102}$}
    \psfrag{d4}[cl][cl][\scale]{BRGC}
%    \psfrag{d5}[cl][cl][\scale]{$q = 5$}
%    \psfrag{d6}[cl][cl][\scale]{$q = 6$}
%    \psfrag{d7}[cl][cl][\scale]{$q = 7$}
%    \psfrag{d8}[cl][cl][\scale]{$q = 8$}
%    \psfrag{d9}[cl][cl][\scale]{$q = 9$}
%    \psfrag{d10}[cl][cl][\scale]{$q = 10$}
    \psfrag{d1111}[cl][cl][\scale]{$\pp_{15}$}
    \includegraphics[width=0.95\columnwidth]{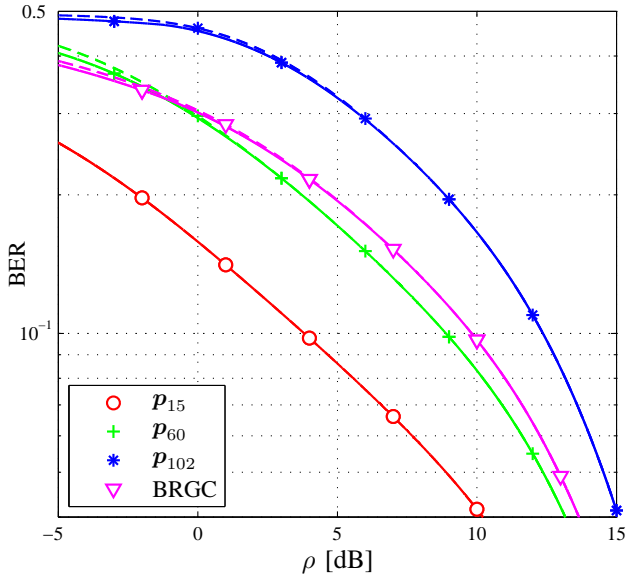}
    \caption{The BER for 8-PAM with patterns $\pp_{15}$, $\pp_{60}$, $\pp_{102}$, and the BRGC. Solid lines correspond to the BD and dashed lines correspond to the ABD.}
    \label{fig:BD_BER}
\end{center}
\end{figure}

To illustrate Remark~\ref{rem:ber_general}, consider 8-PAM labeled by the BRGC, which is formed by the patterns $\pp_{15}= [0,0,0,0,1,1,1,1]$, $\pp_{60} = [0,0,1,1,1,1,0,0]$, and $\pp_{102} = [0,1,1,0,0,1,1,0]$. From~\eqsref{eq:BER_general}{eq:g_ik}, whenever $g_{i,k} = 0$, the value of $\beta_k$ does not influence the PBER and can be set to any value. The thresholds for $g_{i,k} \neq 0$ can be numerically calculated setting the L-value in~\eqref{eq:llr} to zero. The obtained results are shown in~\figref{fig:BD_THR}. Using these thresholds in~\eqref{eq:BER_general} and~\eqref{eq:P_C1.0}, the BER for the patterns and for the BRGC are calculated. The results for the BD and the ABD are presented in~\figref{fig:BD_BER} and show no notable difference between the demodulators for $\SNR > 0$ dB.

%%%%%%%%%%%%%%%%%%%%%%%%%%%%%%%%%%%%%%%%%%%%%%%%%%%%%%%%%%%%%%%%%%%%%%%%%%%%%%%%%%%%%%%%%%%%%%%%%%%%%%%%%%%%%%%%%%%%%%%%%%%%%%%%%%%%%%%%%%%%
%%%%%%%%%%%%%%%%%%%%%%%%%%%%%%%%%%%%%% BER Expression %%%%%%%%%%%%%%%%%%%%%%%%%%%%%%%%%%%%%%%%%%%%%%%%%%%%%%%%%%%%%%%%%%%%%%%%%%%%%%%%%%%%%%
%%%%%%%%%%%%%%%%%%%%%%%%%%%%%%%%%%%%%%%%%%%%%%%%%%%%%%%%%%%%%%%%%%%%%%%%%%%%%%%%%%%%%%%%%%%%%%%%%%%%%%%%%%%%%%%%%%%%%%%%%%%%%%%%%%%%%%%%%%%%

\section{BER for $M$-PAM}\label{sec:BER_analysis}

In this section, we study the BER for equally spaced $M$-PAM constellations. We concentrate on classifying patterns and comparing their performance. For $M$-PAM,~\eqref{eq:BER_general} can be expressed as a bit-wise version of~\cite[eq.~(21)]{Agrell07}:
\begin{equation}
    \Pt = \frac{1}{M}\sum_{n = 1}^{M-1}a_n\qfunc{\left( (2n-1)d\sqrt{2\SNR}\right)},
    \label{eq:BER_maxlog_new}
\end{equation}
where
\begin{multline}
    a_n \triangleq \sum_{k = n}^{M-1} (p_{k+1} - p_{k})(1-2p_{k+1-n}) \\ - (p_{k+2-n} - p_{k+1-n})(1-2p_{k+1}).
        \label{eq:ber_maxlog_weights}
\end{multline}

One direct consequence of~\eqref{eq:BER_maxlog_new} is that the vector $\aa \triangleq [a_1, \dots, a_{M-1}]$ with $a_n$ given by~\eqref{eq:ber_maxlog_weights} completely defines the performance of the ABD for $M$-PAM and allows us to compare the performance of different patterns. From~\eqref{eq:BER_maxlog_new}, the PBER for high SNR can be predicted by the coefficient multiplying the Q-function with the smallest argument, that is, $a_1$. If for two patterns the coefficients are identical, the next coefficients $a_2$ are checked, and so on.

We observe that, for instance, for 4-PAM, the pattern $\pp_5 = [0, 1, 0, 1]$ and the pattern $\pp_{10}= [1, 0, 1, 0]$ have identical PBER performance because of the symmetry of the constellation. It is therefore interesting to find all the patterns with different performance. This will allow us to predict the performance of any possible labeling.
We therefore group all the patterns with identical performance into one class.  The next theorem gives a closed form expression for the number of classes for length-$M$ patterns.
\begin{theorem}\label{the:number_classes}
For $M$-PAM, all the length-$M$ patterns can be grouped into $Q$ classes, where the patterns within each class have identical PBER, and
\begin{align}
    Q &= \frac{1}{4}\left(\tbinom{M}{M/2} + \tbinom{M/2}{M/4}+ 2^{M/2}\right).\label{eq:Q}
\end{align}
\end{theorem}
\begin{IEEEproof}
We define two operations that can be applied to a pattern that will be used in the proof. A \emph{reflection} of $\pp$ is defined as $\pp' = \refl{(\pp)}$ with $p'_i = p_{M +1 - i}$ for $i = 1,\dots,M$. %For example, $\pp_{27} = [0,0,0,1,1,0,1,1]=\refl{([1,1,0,1,1,0,0,0])}=\refl{(\pp_{216})}$.
An \emph{inversion} of $\pp$ is defined as $\pp' = \inv{(\pp)}$ with $p'_i = \bar{p}_i$ for $i = 1,\dots,M$. %For example, $\pp_{39} = [0,0,1,0,0,1,1,1]=\inv{([1,1,0,1,1,0,0,0])}=\inv{(\pp_{216})}$.
Both these functions are self-inverse, i.e., $\pp = \refl(\refl(\pp))$ and $\pp = \inv(\inv(\pp))$, and they commute, i.e., $\refl(\inv(\pp)) = \inv(\refl(\pp))$. Note also that for any pattern $\pp$, we have that $\pp \neq \inv(\pp)$.

We introduce three special types of patterns.
    The pattern $\pp$ is said to be \emph{reflected} (RE) if $\refl({\pp}) = \pp$,
     the pattern $\pp$ is said to be \emph{anti-reflected} (ARE) if $\inv(\refl({\pp})) = \pp$, and
    the pattern $\pp$ is called \emph{asymmetric} (ASY) if it is neither RE nor ARE. For example, $\pp_{60} = [0,0,1,1,1,1,0,0]$ is an RE pattern, $\pp_{43} = [0,0,1,0,1,0,1,1]$ is an ARE pattern, and  $\pp_{216} = [1,1,0,1,1,0,0,0]$ is an ASY pattern.

From \eqsref{eq:P_C1}{eq:P_j}, we note that the PBER is not affected by reflections and/or inversion of the patterns, since the PBER is averaged over both transmitted zeros and ones. Because of this,  we group all patterns that are connected via reflection or inversion into one class of patterns that all have identical PBER. Each class contains either two patterns ($\pp$ and $\inv(\pp)$ because $\pp \neq \inv(\pp) ,\,\forall\pp$) or four patterns ($\pp$, $\inv(\pp)$, $\refl(\pp)$, and $\inv(\refl(\pp))$).% and is represented by a unique class index $q\in {1,\dots, Q}$, where $Q$ is the number of classes.

Any pattern $\pp$ must contain $M/2$ zeros and $M/2$ ones, hence, the total number of patterns is equal to $\tbinom{M}{M/2}$. For a pattern to be RE, $p_i = p_{M-i+1}$, i.e., the positions of the $M/4$ ones in $[p_1,\dots,p_{M/2}]$ fully describe the pattern, and thus, the number of RE patterns is $\tbinom{M/2}{M/4}$. There are two members in every class of RE patterns, $\pp = \refl(\pp)$ and $\inv(\refl(\pp)) = \inv(\pp)$, which gives~$\frac{1}{2}\tbinom{M/2}{M/4}$ classes.

For a pattern to be ARE, $p_i = \bar{p}_{M-i+1}$, i.e., the positions of the ones in $[p_1,\dots,p_{M/2}]$ fully describe the pattern, where the number of ones in $[p_1,\dots,p_{M/2}]$ is between 0 and $M/2$. From that, it follows that there are $2^{M/2}$ ARE patterns. There are two members in every class of ARE patterns ($\pp = \inv(\refl(\pp))$ and $\refl(\pp) = \inv(\pp)$), which gives~$2^{M/2-1}$ classes.

All the remaining classes include only ASY patterns. The number of ASY patterns can be obtained by subtracting the number of RE and ARE patterns from the total number of patterns. There are four patterns in each class, as  $\pp \neq \refl(\pp)$ and $\pp \neq  \refl(\inv(\pp))$ (or equivalently, $ \refl(\pp) \neq  \inv(\pp)$). Using this, the total number of classes in~\eqref{eq:Q} is obtained as sum of classes of RE, ARE, and ASY patterns.
\end{IEEEproof}

\begin{figure}[t]
\newcommand{\scale}{0.9}
\begin{center}
    \psfrag{SNR}[tc][tc]{$\SNR$~[dB]}
    \psfrag{BER}[bc][Bc][\scale]{PBER}
    \psfrag{a1}[r][c][\scale]{$a_1 = 2$}
    \psfrag{a2}[r][c][\scale]{$a_1 = 4$}
    \psfrag{a3}[r][c][\scale]{$a_1 = 6$}
    \psfrag{a4}[l][c][\scale]{$a_1 = 8$}
    \psfrag{a5}[l][c][\scale]{$a_1 = 10$}
    \psfrag{a6}[l][c][\scale]{$a_1 = 12$}
    \psfrag{a7}[l][c][\scale]{$a_1 = 14$}
    \psfrag{a8}[r][c][\scale]{$a_1 = 2$}
    \psfrag{a9}[c][B][\scale]{$a_1 = 4$}
    \psfrag{a10}[c][t][\scale]{$a_1 = 30$}
    \subfigure[8-PAM]{\includegraphics[width=0.95\columnwidth]{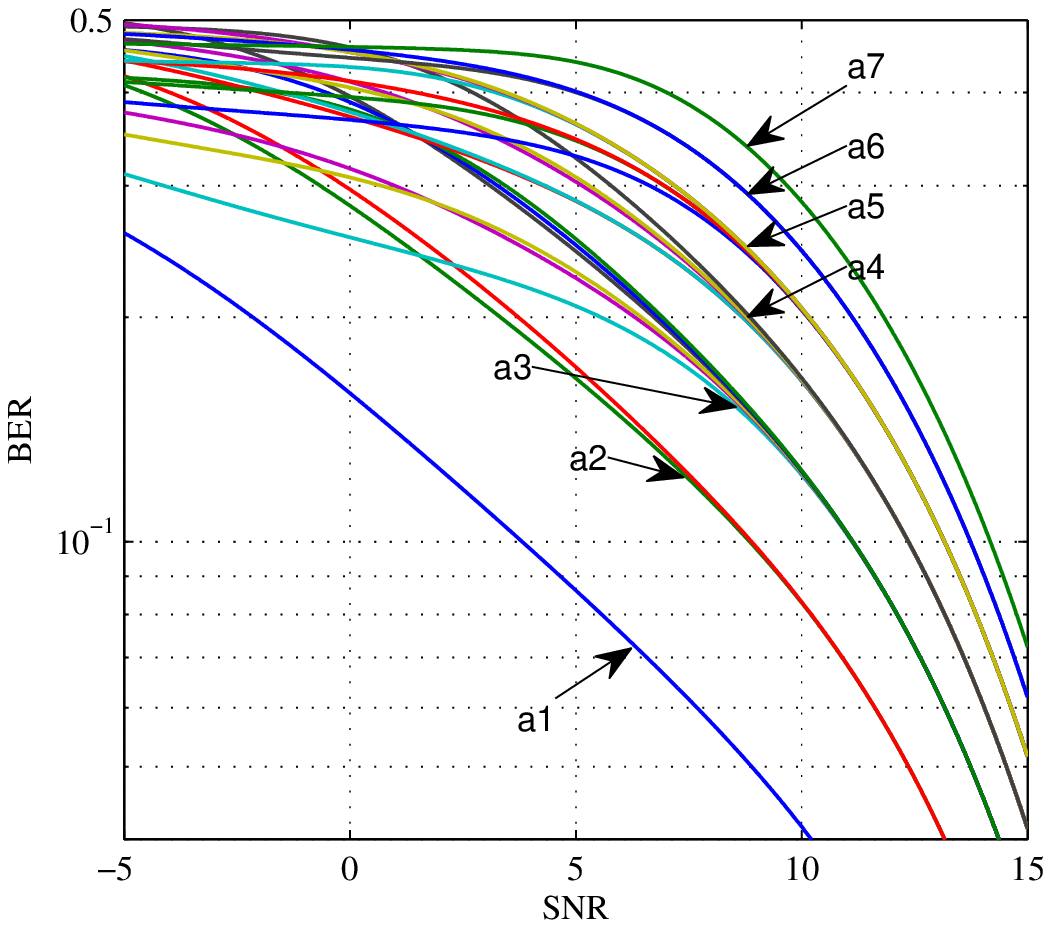}}
    \subfigure[16-PAM]{\includegraphics[width=0.95\columnwidth]{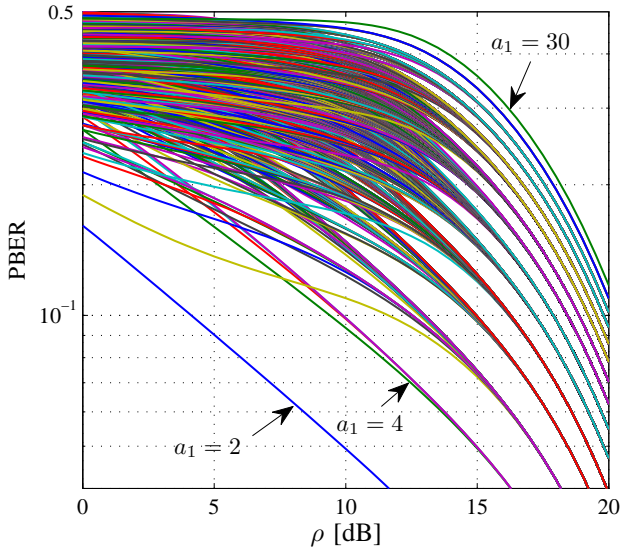}}
    \caption{The PBER for the patterns for 8-PAM and 16-PAM. All the curves merge into $M-1$ groups at high SNR as predicted by Remark~\ref{rem:high_snr_pattern}.}
    \label{fig:BER4ABD}
\end{center}
\end{figure}

For example, \theref{the:number_classes} states that there are 3 classes of patterns for 4-PAM, 23 classes for 8-PAM, and 3299 classes for 16-PAM. The PBER for 8-PAM and 16-PAM for all the patterns is shown in~\figref{fig:BER4ABD}. All the classes of patterns for 4-PAM and 8-PAM are shown in the first and the second parts of~\tabref{tab:all_seq}, respectively. For each class,~\tabref{tab:all_seq} shows the representative of the class $\pp$, the decimal indices of class members $w$ (the index of the representative is shown with boldface), and the vector $\aa$ that defines the PBER. The patterns are ordered from best to worst PBER at high SNR.

\begin{table}[t]
    \renewcommand{\arraystretch}{1.0}
\footnotesize
    \centering
        \caption{Classes of patterns for 4-PAM and $8$-PAM with their corresponding representatives $\pp$ , decimal representations of the patterns $w$, and vectors $\aa$ defining their PBER}
        \begin{tabular}{@{~}c@{~}|@{~}c@{~}|@{~}c@{~}}
            \hline
            $\pp$ &$w$& $\aa$\\
            \hline

            \hline
            $[0,0,1,1]$&${\bf3} \tab 12$ & $[2, \tab\tabb2, \tabb\tab0]$\\
            \hline
            $[0,1,1,0]$&${\bf6} \tab\tab 9$ & $[4, \tabb\tab2, -2]$ \\
            \hline
            $[0,1,0,1]$&${\bf5}\tab 10 $& $[6, -4, \tab\tabb2]$ \\
            \hline
            \hline
            $[0,0,0,0,1,1,1,1]$&$\bf{15}$ $240$&$[\tab2,\tab 2,\tab\, 2,\tab 2,\tab 0,\,\tab 0,\tab 0]$\\
            \hline
            $[0,0,0,1,1,1,1,0]$& ${\bf30}$ $120$ $135$ $225$& $[\tab4,\tabb 3,\tabb 3,\tabb 2, -2, -1, -1]$\\
            \hline
            $[0,0,1,1,1,1,0,0]$&${\bf60}$ $195$ &$[\tab4,\tab4,\tab2,\tabb2,-2,-2,\tab0]$\\
            \hline
            $[0,0,0,1,0,1,1,1]$&$\bf{23}$  $232$  &$[\tab6, -2,\tab 2,\tab 0,\tab 2,\tab 0,\tab 0]$  \\
            \hline
            $[0,0,0,1,1,1,0,1]$& ${\bf29}$ \hspace{0.5mm} $71$ $184$ $226$& $[\tab6, \tab 1, \tab2, -3, \tab 1,\tab 0,\tab 1]$\\
            \hline
            $[0,0,0,1,1,0,1,1]$& ${\bf27}$\hspace{1.5mm} $39$ $216$ $228$ & $[\tab6, \tab 2, -3, \tab1,\tab 1, \tab1,\tab 0]$\\
            \hline
            $[0,1,1,1,0,0,0,1]$& ${\bf113}$ $142$ & $[\tab6, \tabb4,\tabb 4, -4, -2, -2,\tabb 2]$\\
            \hline
            $[0,0,1,1,1,0,0,1]$& ${\bf57}$ \hspace{0.5mm} $99$ $156$ $198$ & $[\tab6, \tab5, \tabb 0, -3, -3, \tab2, \tab1]$\\
            \hline
            $[0,0,1,1,0,0,1,1]$&${\bf51}$ $204$ & $[\tab6,\tab6,-4,-4,\tab2,\tabb2,\tab0]$\\
            \hline
            $[0,0,1,0,1,1,1,0]$& ${\bf46}$ $116$ $139$ $209$& $[\tab8, -1, 2, -1,\, 3, -2, -1]$ \\
            \hline
            $[0,0,1,1,1,0,1,0]$& ${\bf58}$ \hspace{0.5mm} $92$ $163$ $197$ & $[\tab8, -1, 3, -2,\, 2, -1, -1]$\\
            \hline
            $[0,1,0,0,1,1,1,0]$& ${\bf78}$ $114$ $141$ $177$& $[\tab8, \,2, -1, -1, -1, 3, -2]$\\
            \hline
            $[0,0,1,1,0,1,1,0]$& ${\bf54}$ $108$ $147$ $201$ & $[\tab8, \tabb3, -6,\tabb 3,\tabb 3, -2, -1]$\\
            \hline
            $[0,1,1,0,0,1,1,0]$&${\bf102}$ $153$ & $[\tab8,\tabb 6, -6, -4,\tabb 4,\tabb 2, -2]$\\
            \hline
            $[0,0,1,0,1,0,1,1]$&$\bf{43}$ $212$& $[10, -6,\tab 4, -2,\tabb 0,\tab 2,\tab 0]$ \\
            \hline
            $[0,0,1,0,1,1,0,1]$& ${\bf45}$ \hspace{0.5mm} $75$ $180$ $210$& $[10, -3, -3, \tabb6, -4,\tabb 1,\tabb 1]$\\
            \hline
            $[0,0,1,1,0,1,0,1]$& ${\bf53}$ \hspace{0.5mm} $83$ $172$ $202$  & $[10, -3, \tab1, \tab0, -2, \tab1,\tabb 1]$\\
            \hline
            $[0,1,0,0,1,1,0,1]$&${\bf77}$ $178$ & $[10,\tabb 0, -6,\tab 2,\tab 4, -4,\tab 2]$ \\
            \hline
            $[0,1,1,0,1,0,0,1]$& ${\bf105}$ $150$& $[10, \tabb 0, -4, \tabb 6, -4, -2, \tabb 2]$\\
            \hline
            $[0,1,0,1,1,0,0,1]$& ${\bf89}$ $101$ $154$ $166$ &$[10, \tab0, -3, \tab1,\tab 1, -3,\tabb 2]$\\
            \hline
            $[0,1,0,1,1,0,1,0]$&$\bf{90}$ $165$ & $[12, -6,\tabb 0,\tabb 6, -6, \tabb 4, -2]$\\
            \hline
            $[0,1,0,1,0,1,1,0]$& ${\bf86}$ $106$ $149$ $169$& $[12, -6, 3, -1, -1, \,3, -2]$\\
            \hline
            $[0,1,0,1,0,1,0,1]$&${\bf85}$ $170$ & $[14, -12, 10, -8, 6, -4, 2]$\\
            \hline

            \hline
        \end{tabular}
        \label{tab:all_seq}
\end{table}

\begin{remark}\label{rem:high_snr_pattern}
The element $a_1$ in~\eqref{eq:ber_maxlog_weights} is equal to twice the number of pairs of constellation points at minimum ED whose bits are different (for a given pattern). Using this, it can be shown that for $M$-PAM there are $M-1$ different values of $a_1$. This means that the PBER of all the patterns merge into $M-1$ groups at high SNR. For example, for 8-PAM and 16-PAM the number of groups of patterns at high SNR is 7 and 15, respectively, as illustrated by~\figref{fig:BER4ABD}.
\end{remark}

Using~\eqref{eq:P_C1.0} and \eqref{eq:BER_maxlog_new}, the average BER for $M$-PAM with labeling $\C$ can be expressed as~\cite[eq.~(21)]{Agrell07}:
\begin{equation}
    \PC = \frac{1}{mM}\sum_{n = 1}^{M-1}\alpha_n\qfunc{\left( (2n-1)d\sqrt{2\SNR}\right)},
\end{equation}
where $\aalpha \triangleq [\alpha_1, \dots, \alpha_{M-1}]$ is the sum of vectors $\aa$ for the $m$ patterns used in $\C$. The value of $\alpha_n$ is a scaled version of the so-called differential average distance spectrum $\bar{\delta}(n,\lambda)$ in~\cite[eq.~21]{Agrell07}, i.e., $\alpha_n = 2M\bar{\delta}(n, \lambda)$.

\begin{remark}
The value of $\alpha_1$ corresponds to twice the sum of Hamming distances between binary labelings of constellation points at minimum ED. It can be shown that $A_{\phi} = 2m(M-1) - \alpha_1$, where $A_{\phi}$ was recently shown to determine the BICM mutual information in the high SNR regime \cite{Alvarado11oct}.
\end{remark}

\begin{figure}[t]
\newcommand{\scale}{0.9}
\begin{center}
    \psfrag{SNR}[tc][tc]{$\SNR$~[dB]}
    \psfrag{BER}[bc][Bc][\scale]{BER}
    \psfrag{AG}[l][c][\scale]{AG}
    \psfrag{BRGC}[r][c][\scale]{BRGC}
    \psfrag{NBC}[r][c][\scale]{NBC}
    \psfrag{FBC}[r][c][\scale]{FBC}
    \psfrag{BSGC}[r][c][\scale]{\hspace{-15mm}BSGC}
    \includegraphics[width=0.95\columnwidth]{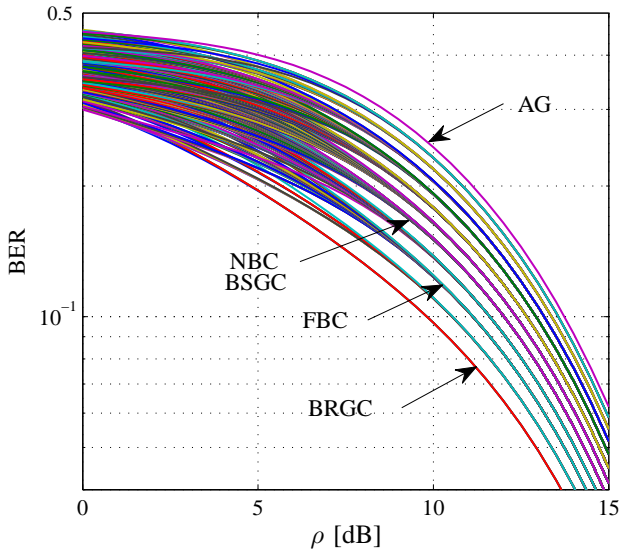}
    \caption{The BER for all the 460 labelings with different BER for 8-PAM.}
    \label{fig:PAM8lab}
\end{center}
\end{figure}

By listing the vectors $\aalpha$ for all the possible labelings for 8-PAM, we found 12 different $\alpha_1$, which is in agreement with the 12 classes of labelings (with different $A_{\phi}$) shown in~\cite[Fig.~2(b)]{Alvarado11oct}. The  BER for all the labelings for 8-PAM is shown in~\figref{fig:PAM8lab}, where the 12 classes are visible for high SNR.

\begin{table}[t]
    \renewcommand{\arraystretch}{1.2}
    \centering
    \footnotesize
    \caption{Some common labelings for $4$-PAM and $8$-PAM with their corresponding pattern indices $\setW$ and vectors $\aalpha$ defining their BER }
        \begin{tabular}{@{~}c@{~}|@{~}c@{~}|@{~}c@{~}|@{~}l@{~}}
            \hline
            $M$ & Labeling &$\setW$ & $\aalpha$ \\
            \hline

            \hline
            $4$ &BRGC  & $\{3, 6\}$  & $[6, 4, -2]$ \\
            \hline
            $4$ & NBC   & $\{3, 5\}$  & $[8, -2, 2]$ \\
            \hline
            $4$ & AG    & $\{5, 6\}$  & $[10, -2, 0]$ \\
            \hline
            \hline
            $8$ & BRGC  & $\{15, 60, 102\}$  & $[14, 12, -2, 0, 2, 0, -2]$ \\
            \hline
            $8$ & FBC  & $\{15, 60, 90\}$   & $[18, 0, 4, 10, -8, 2, -2]$ \\
            \hline
            $8$ & NBC  & $\{15, 51, 85\}$   & $[22, -4, 8, -10, 8, -2, 2]$ \\
            \hline
            $8$ & BSGC & $\{105, 60, 102\}$ & $[22, 10, -8, 4, -2, -2, 0]$ \\
            \hline
            $8$ & AG   & $\{90, 105, 85\}$  & $[36, -18, 6, 4, -4, -2, 2]$ \\
            \hline

            \hline
        \end{tabular}
        \label{tab:MPAM_labelings}
\end{table}

To conclude, we present the vectors $\aalpha$ for 4-PAM and 8-PAM with some common labelings, including the BRGC, the natural binary labeling (NBC)~\cite[Sec.~II-B]{Agrell11}, the folded binary code (FBC)~\cite{Lassing03b}~\cite[Sec.~II-B]{Agrell11}, the binary semi-Gray code (BSGC)~\cite[Sec.~II-B]{Agrell11}, and the so-called anti-Gray (AG) labeling~\cite{Brink98b}. These labelings are shown in~\tabref{tab:MPAM_labelings} together their pattern indices $\setW$ and vectors $\aalpha$, in the first part for 4-PAM, and in the second part for 8-PAM. The labelings are also ordered from best to worst BER at high SNR. By listing the vectors $\aalpha$ for all the possible labelings, we found three labelings with different BER for 4-PAM listed in~\tabref{tab:MPAM_labelings}. For 8-PAM we found 460 labelings as shown in~\figref{fig:PAM8lab}.

%%%%%%%%%%%%%%%%%%%%%%%%%%%%%%%%%%%%%%%%%%%%%%%%%%%%%%%%%%%%%%%%%%%%%%%%%%%%%%%%%%%%%%%%%%%%%%%%%%%%%%%%%%%%%%%%%%%%%%%%%%%%%%%%%%%%%%%%%%%%
%%%%%%%%%%%%%%%%%%%%%%%%%%%%%%%%%%%%%% Conclusion %%%%%%%%%%%%%%%%%%%%%%%%%%%%%%%%%%%%%%%%%%%%%%%%%%%%%%%%%%%%%%%%%%%%%%%%%%%%%%%%%%%%%%%%%%
%%%%%%%%%%%%%%%%%%%%%%%%%%%%%%%%%%%%%%%%%%%%%%%%%%%%%%%%%%%%%%%%%%%%%%%%%%%%%%%%%%%%%%%%%%%%%%%%%%%%%%%%%%%%%%%%%%%%%%%%%%%%%%%%%%%%%%%%%%%%
\section{Conclusions} \label{sec:conclusions}

A novel general expression for the uncoded BER of one-dimensional constellations has been introduced. For equally spaced $M$-PAM constellations, a classification of the patterns has been performed and a closed form expression on the number of patterns that give different BER has been derived. The rule for combining patterns into a labeling remains for future investigation. Establishing this rule will allow us to define a number of labelings with different BER for an arbitrary constellation size $M$ and also a number of groups of labelings at high SNR.

\balance

\bibliographystyle{IEEEtran}
\bibliography{./bibliography/MyBibliography}
\end{document}